# Global Trigger Technological Demonstrator for ATLAS Phase-II upgrade

Viacheslav Filimonov, Bruno Bauss, Volker Büscher, Ulrich Schäfer, Duc Bao Ta

***Abstract*—ATLAS detector at the LHC will undergo a major Phase-II upgrade for the High Luminosity LHC. The upgrade affects all major ATLAS systems, including the Trigger and Data Acquisition systems.**

**As part of the Level-0 Trigger System, the Global Trigger uses full-granularity calorimeter cells to perform algorithms, refines the trigger objects and applies topological requirements. The Global Trigger uses a Global Common Module as the building block of its design. To achieve a high input and output bandwidth and substantial processing power, the Global Common Module will host the most advanced FPGAs and optical modules.**

**In order to evaluate the new generation of optical modules and FPGAs running at high data rates (up to 28 Gb/s), a Global Trigger Technological Demonstrator board has been designed and tested.**

**The main hardware blocks of the board are the Xilinx Virtex Ultrascale+ 9P FPGA and a number of optical modules, including high-speed Finisar BOA and Samtec FireFly modules. Long-run link tests have been performed for the Finisar BOA and Samtec FireFly optical modules running at 25.65 and 27.58 Gb/s respectively. Successful results demonstrating a good performance of the optical modules when communicating with the FPGA have been obtained.**

**The paper provides a hardware overview and measurement results of the Technological Demonstrator.**

***Index Terms*— LHC, ATLAS, FPGA, Optical Modules**

## I. Introduction

A Global Trigger System is a part of the Phase-II upgrade of the ATLAS [1] Trigger and Data Acquisition (TDAQ) system [2]. The Global Trigger will replace the Phase-I Topological Processor [3] and extend its functions by using full-granularity calorimeter cells for refining the trigger objects calculated by the Level-0 Trigger System, performing offline-like algorithms, including iterative algorithms such as topoclustering and the use of Higher Level Synthesis, calculating event-level quantities and applying topological requirements (Figure 1).

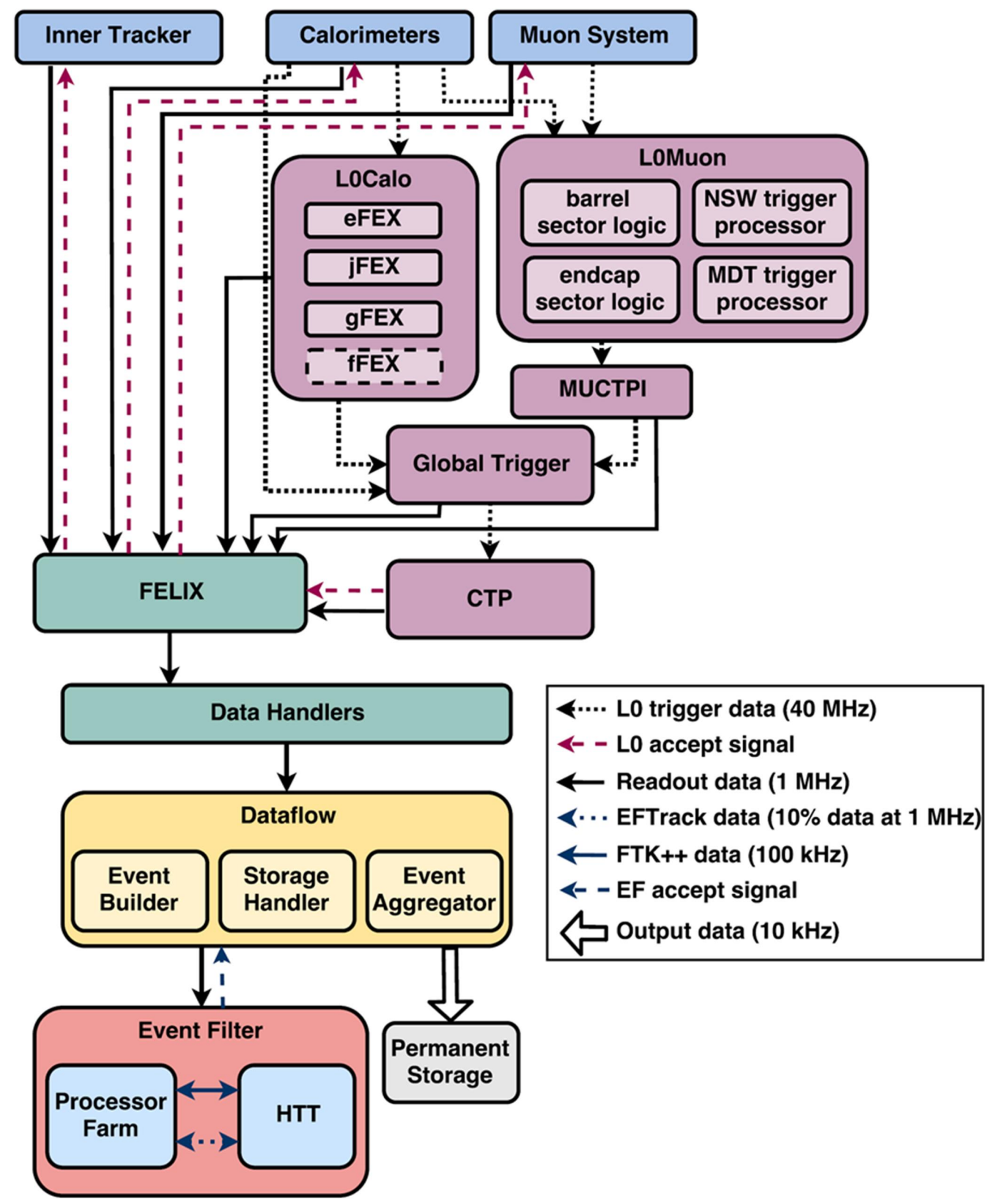

Figure 1. TDAQ System after the Phase-II upgrade (baseline design) [2].

The Global Trigger is a time-multiplexed system, which concentrates data of full event into a single processor. The Global Trigger System is composed of three main layers: Multiplexing (MUX) layer, Global Event Processor (GEP) layer and Demultiplexing Global-to-Central Trigger Processor (CTP) Interface (Figure 2).

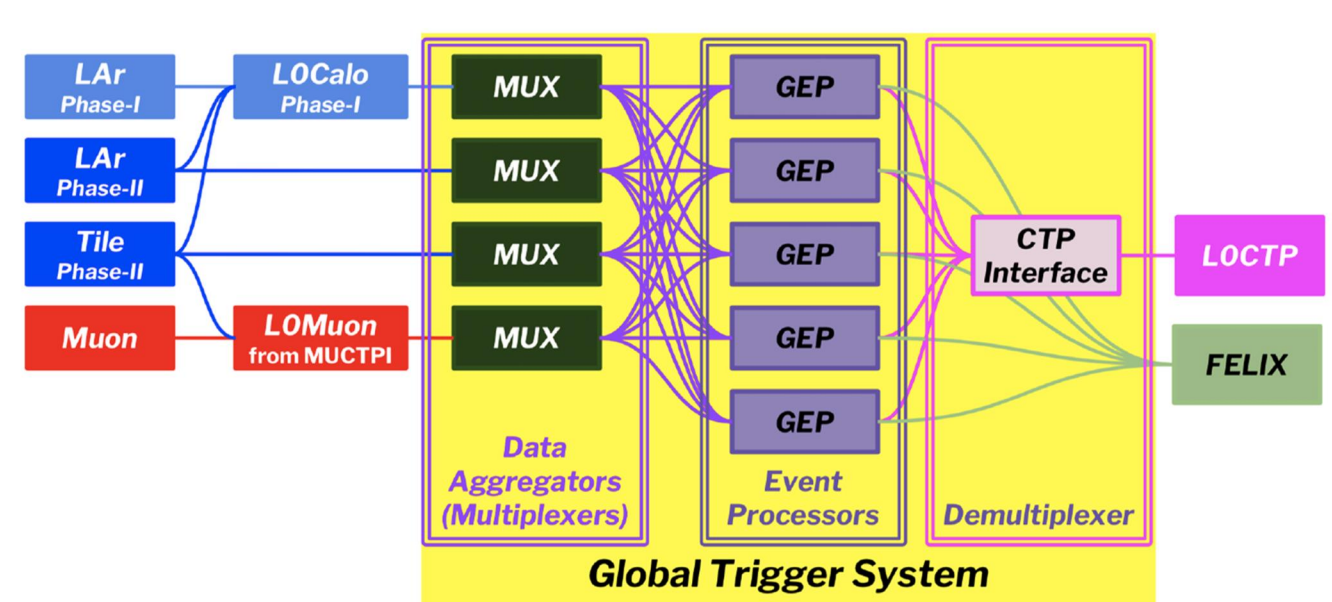

Figure 2. Schematic view of the Global Trigger System [2].

All authors are with Institut für Physik, Johannes Gutenberg Universität Mainz, Germany. Corresponding author is Viacheslav Filimonov (email: viacheslav.filimonov@cern.ch).

The building block of the Global Trigger is the Global Common Module (GCM), which has to provide significant processing resources along with a high input and output bandwidth. This ATCA module is used in each layer of the Global Trigger System.

A block diagram for the GCM is shown in Figure 3. The board hosts two large FPGAs (Xilinx Ultrascale+ VU13P [5]) with a sufficient number of multi-gigabit transceivers. A central processing chip (ZYNQ MPSoC) is implemented for monitoring, control and readout functions. Random Access Memory (RAM) storage modules are used for long latency buffering of the input and output data. Each of the large FPGAs serves as a MUX or a GEP node, as shown in Figure 2.

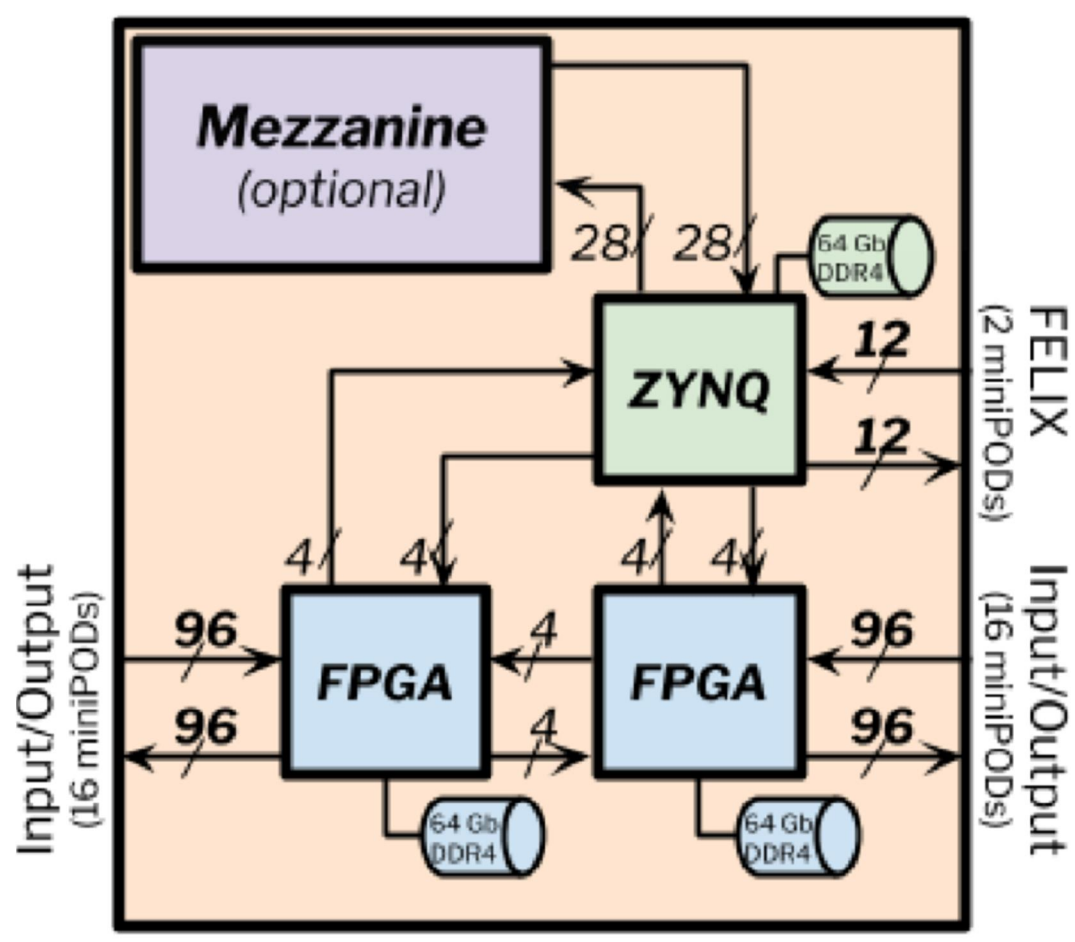

Figure 3. GCM block diagram [2].

In order to provide a platform to evaluate various firmware blocks (MUX and GEP) of the Global Trigger System during the development of the GCM infrastructure and algorithm firmware, the Production Firmware Deployment Module (PFM) is used. This board represents a slice of the GCM, which includes a single processing unit (MUX or GEP), a control FPGA and a number of optical modules.

The PFM is designed in an ATCA form factor with the possibility of a standalone operation. The structure of the board can be seen in Figure 4. The layout is well-matched to the ATCA airflow.

The main building blocks are the following: single MUX / GEP FPGA (Xilinx Ultrascale+ VU13P), up to 8 Finisar BOA [7] modules for real-time data path, single Finisar BOA module for interface to Front-End Link eXchange (FELIX) system, UltraZed [8] board with Zynq UltraScale+, IPM Controller (IPMC), power mezzanines and DDR4 RAMs.

Such a configuration of the PFM allows testing and debugging algorithm and infrastructure firmware together with the corresponding software, thus minimizing the risk of firmware failures on the final system. In order to allow for early firmware development and evaluation, the PFM needs to be available well before the production GCM.

The design of the PFM is based on the R&D performed with the Technological Demonstrator, described below.

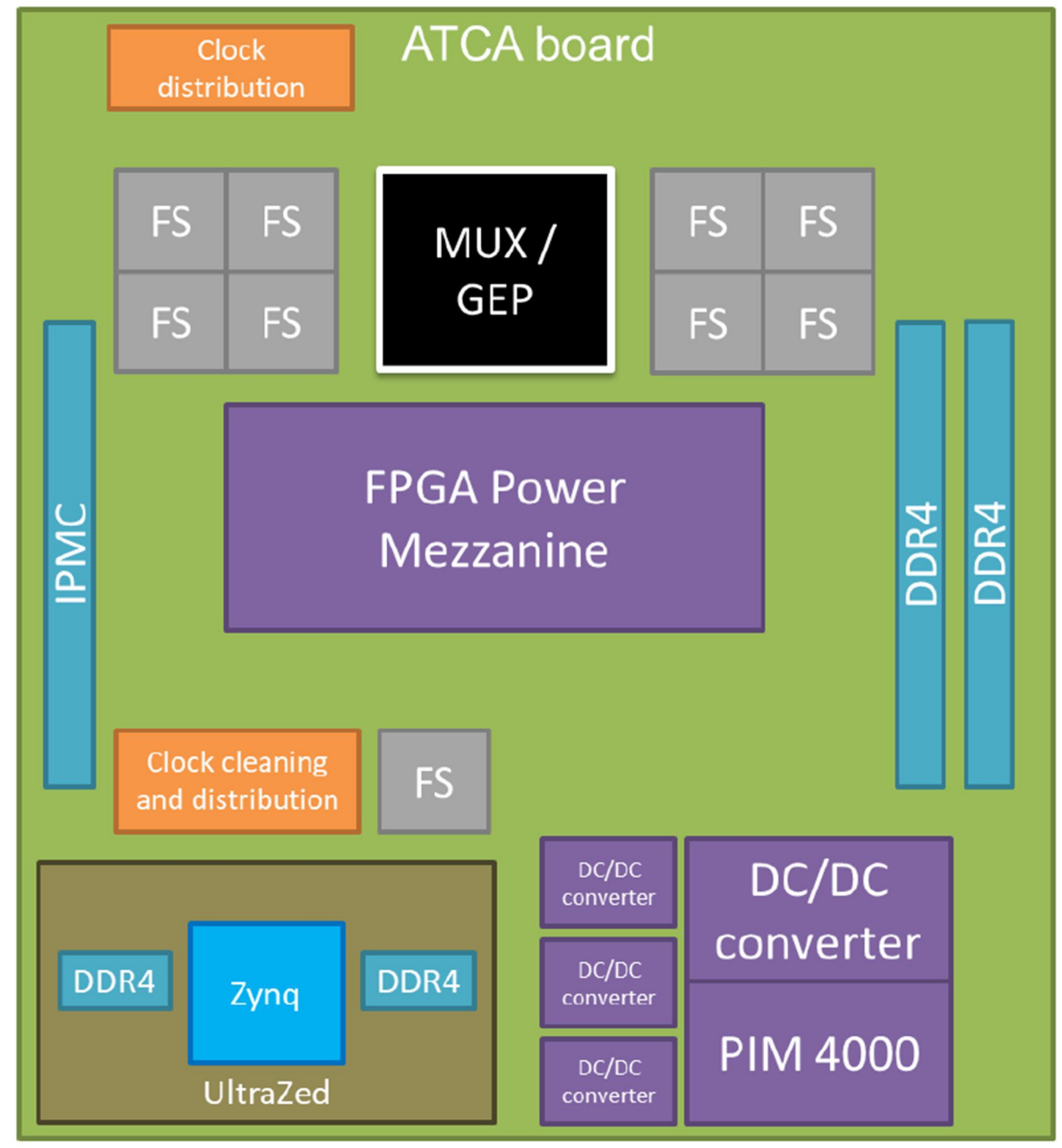

Figure 4. PFM hardware structure.

Various link speeds, including high-speed links up to 28 Gb/s, should be supported on the GCM and the PFM. The high-speed link support is essential in order to cope with the transmission of high-granularity calorimeter data, which drives the bandwidth requirement for the upgraded TDAQ system. Since Avago MiniPOD [4] optical modules, widely used in the Phase-I Level-1 Trigger system, can only achieve data rates of 14 Gb/s, it is necessary to evaluate the new generation of optical modules running at high data rates (up to 28 Gb/s). The optical modules should be tested together with the one of the most advanced FPGA supporting high-speed data transmission and having the substantial processing power.

## II. Board Overview

As part of the R&D for the Phase-II Global Trigger System, a Technological Demonstrator board has been designed with an FPGA and high-speed optical modules implemented on-board (Figure 5). The board design is significantly closer to the final GCM and PFM designs in comparison with commercial evaluation boards and, therefore, can provide more reliable evaluation results.

The Demonstrator is designed in a custom ATCA form factor with a number of design blocks, which can be evaluated and reused for the GCM. The central part of the board is the Xilinx Virtex Ultrascale+ 9P FPGA [5], which is connected to two 28G 2x4-lane bidirectional Samtec FireFly [6] modules, one 28G 2x12 bidirectional Finisar BOA [7] module and six MiniPODs. The Demonstrator makes use of power and control mezzanines in order to provide the necessary voltages for the on-board components as well as to provide the essential control functionality. Dedicated clock distribution circuits are implemented as well in order to provide reference clocks for the multi-gigabit transceivers of the FPGA. A slot for an IPMC module, a standard component in ATCA boards for hot-swap power management and sensor monitoring, is present on the board as well.

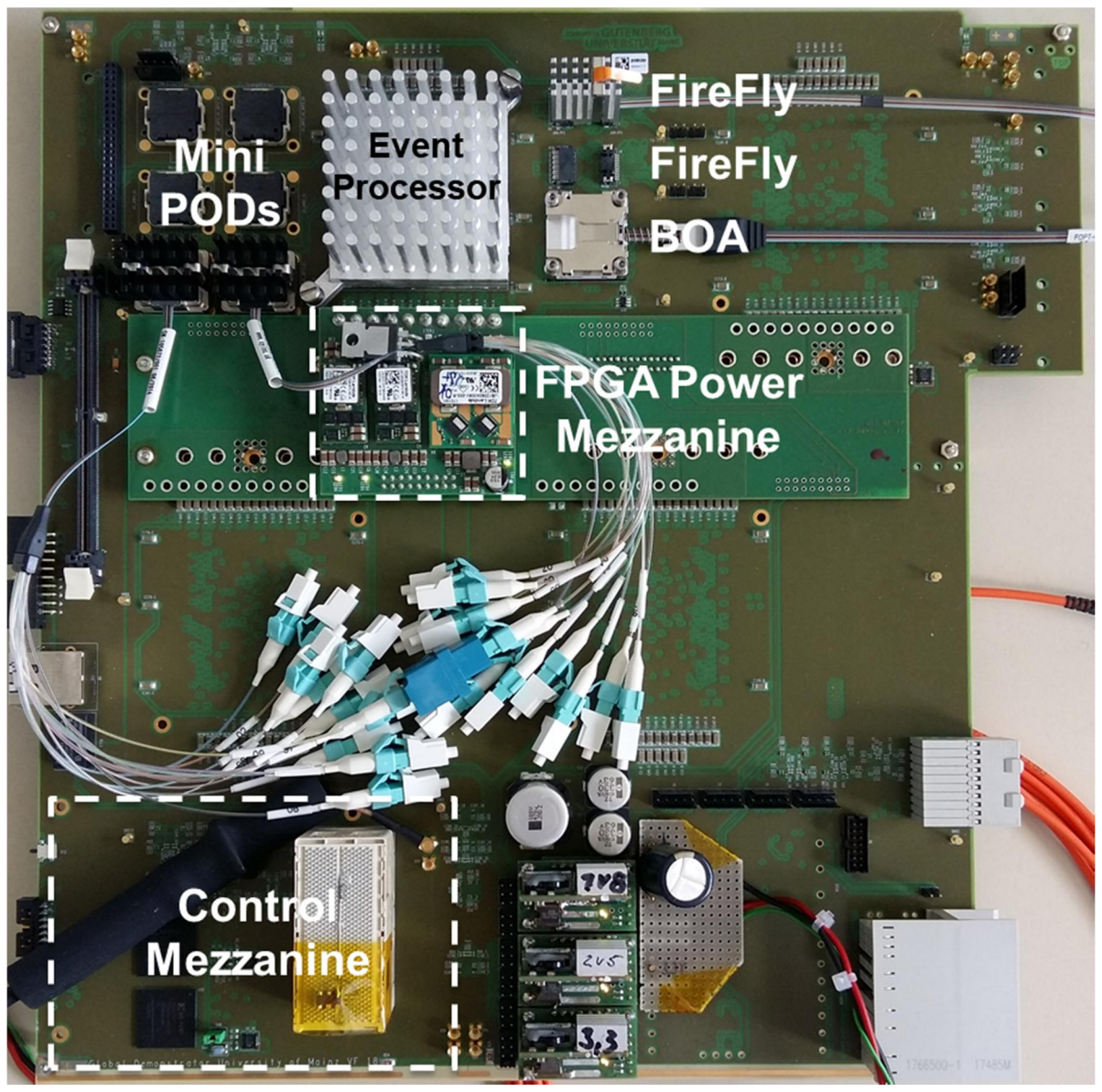

Figure 5. Global Trigger Technological Demonstrator hardware overview.

Such a custom design of the Demonstrator allows evaluating the optical modules together with the FPGA operating on the same mainboard.

## III. Performance Evaluation

Performance of the high-speed optical modules and the FPGA has been evaluated with long-run link tests.

An Integrated Bit Error Ratio Test (IBERT) loopback test has been performed for the Finisar BOA optical module. In the test 12 transmitter links of the optical module were looped back to 12 receiver links of the same module with a help of a 24 to 2x12-fiber Y-cable and a 12-fiber trunk cable (Figure 6).

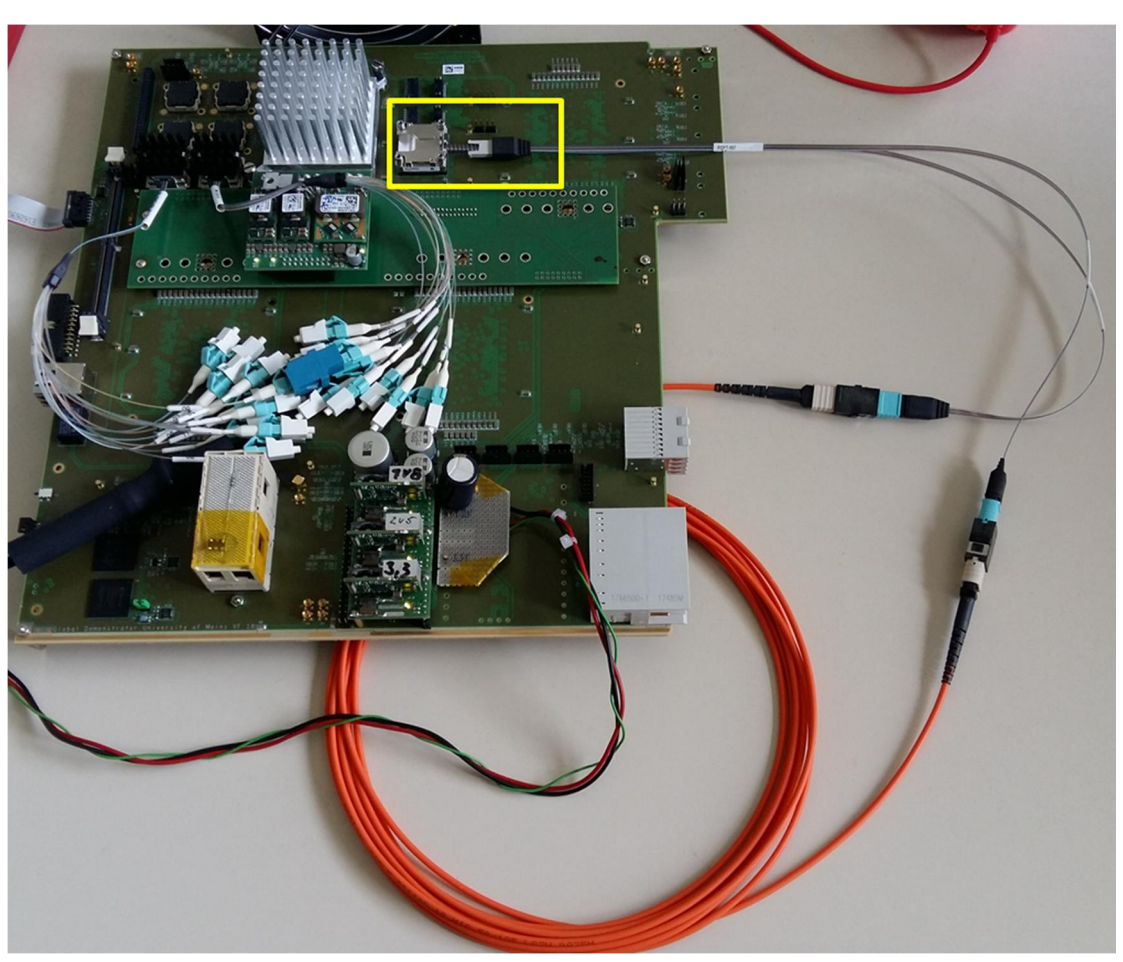

Figure 6. Finisar BOA IBERT loopback test: test setup.

During a day-long IBERT test run at 25.65 Gb/s, using a 31-bit PRBS pattern, a 1.9E-15 BER has been reached. All 12 links are functional, and no bit errors have been detected (Figure 7).

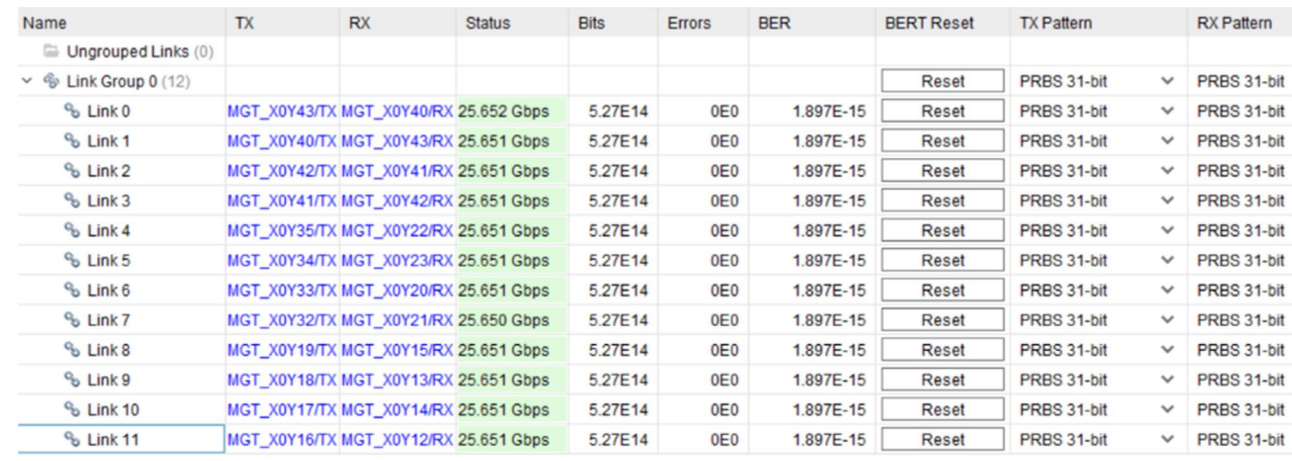

| Name | TX | RX | Status | Bits | Errors | BER | BERT Reset | TX Pattern | RX Pattern |
|---|---|---|---|---|---|---|---|---|---|
| Ungrouped Links (0) | | | | | | | | | |
| Link Group 0 (12) | | | | | | | Reset | PRBS 31-bit | PRBS 31-bit |
| Link 0 | MGT_X0Y43/TX | MGT_X0Y40/RX | 25.652 Gbps | 5.27E14 | 0E0 | 1.897E-15 | Reset | PRBS 31-bit | PRBS 31-bit |
| Link 1 | MGT_X0Y40/TX | MGT_X0Y43/RX | 25.651 Gbps | 5.27E14 | 0E0 | 1.897E-15 | Reset | PRBS 31-bit | PRBS 31-bit |
| Link 2 | MGT_X0Y42/TX | MGT_X0Y41/RX | 25.651 Gbps | 5.27E14 | 0E0 | 1.897E-15 | Reset | PRBS 31-bit | PRBS 31-bit |
| Link 3 | MGT_X0Y41/TX | MGT_X0Y42/RX | 25.651 Gbps | 5.27E14 | 0E0 | 1.897E-15 | Reset | PRBS 31-bit | PRBS 31-bit |
| Link 4 | MGT_X0Y35/TX | MGT_X0Y22/RX | 25.651 Gbps | 5.27E14 | 0E0 | 1.897E-15 | Reset | PRBS 31-bit | PRBS 31-bit |
| Link 5 | MGT_X0Y34/TX | MGT_X0Y23/RX | 25.651 Gbps | 5.27E14 | 0E0 | 1.897E-15 | Reset | PRBS 31-bit | PRBS 31-bit |
| Link 6 | MGT_X0Y33/TX | MGT_X0Y20/RX | 25.651 Gbps | 5.27E14 | 0E0 | 1.897E-15 | Reset | PRBS 31-bit | PRBS 31-bit |
| Link 7 | MGT_X0Y32/TX | MGT_X0Y21/RX | 25.650 Gbps | 5.27E14 | 0E0 | 1.897E-15 | Reset | PRBS 31-bit | PRBS 31-bit |
| Link 8 | MGT_X0Y19/TX | MGT_X0Y15/RX | 25.651 Gbps | 5.27E14 | 0E0 | 1.897E-15 | Reset | PRBS 31-bit | PRBS 31-bit |
| Link 9 | MGT_X0Y18/TX | MGT_X0Y13/RX | 25.651 Gbps | 5.27E14 | 0E0 | 1.897E-15 | Reset | PRBS 31-bit | PRBS 31-bit |
| Link 10 | MGT_X0Y17/TX | MGT_X0Y14/RX | 25.651 Gbps | 5.27E14 | 0E0 | 1.897E-15 | Reset | PRBS 31-bit | PRBS 31-bit |
| Link 11 | MGT_X0Y16/TX | MGT_X0Y12/RX | 25.651 Gbps | 5.27E14 | 0E0 | 1.897E-15 | Reset | PRBS 31-bit | PRBS 31-bit |

Figure 7. Finisar BOA IBERT loopback test: links status.

A typical eye diagram, obtained using a low power mode of the GTY receiver, is shown in Figure 8. With an open area and an open UI of 6656 and 44.44 % respectively, a good performance of the Finisar BOA optical module is achieved.

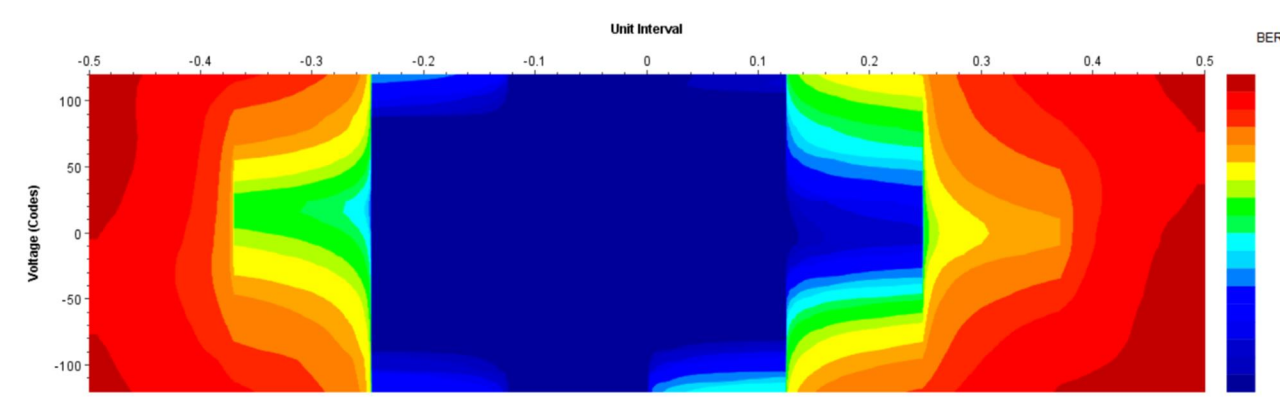

Figure 8. Finisar BOA IBERT loopback test: a typical eye diagram.

An IBERT loopback test has been performed for the Samtec FireFly optical module as well. In the test 4 transmitter links of the optical module were looped back to 4 receiver links of the same module with a help of a 12-fiber MTP to LC breakout cable (Figure 9).

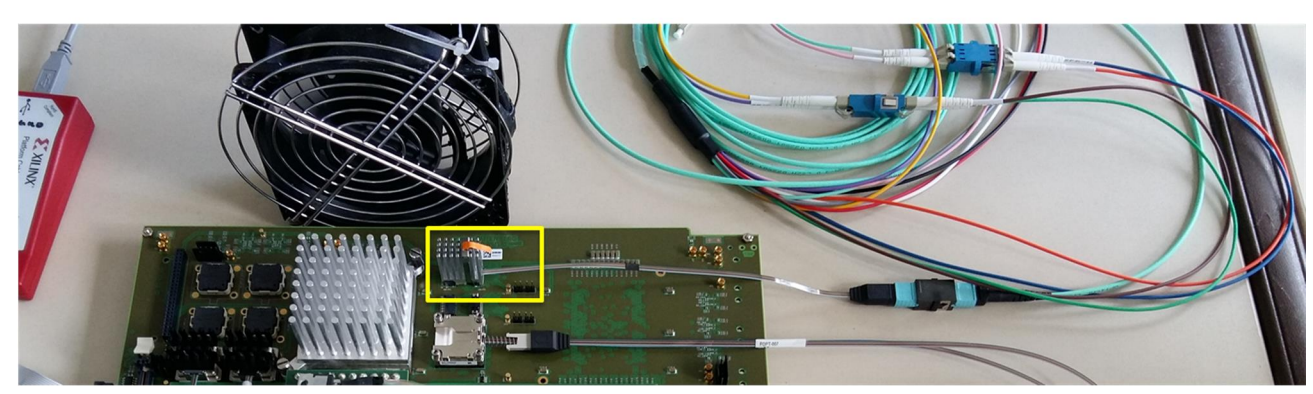

Figure 9. Samtec FireFly IBERT loopback test: test setup.

During a day-long IBERT test run at 27.58 Gb/s, using a 31-bit PRBS pattern, a 1.7E-15 BER has been reached. All 4 links are functional, and no bit errors have been detected (Figure 10).

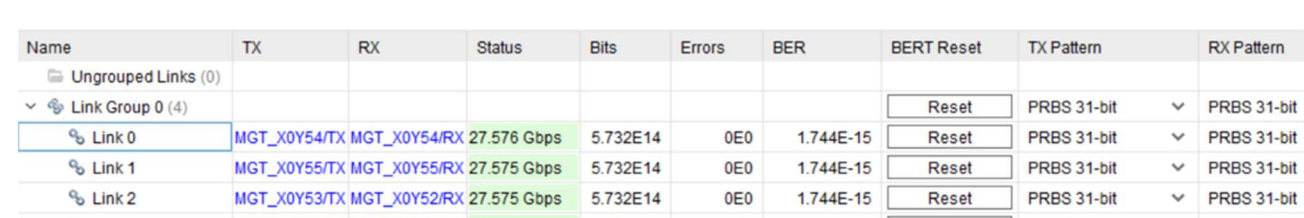

| Name | TX | RX | Status | Bits | Errors | BER | BERT Reset | TX Pattern | RX Pattern |
|---|---|---|---|---|---|---|---|---|---|
| Ungrouped Links (0) | | | | | | | | | |
| Link Group 0 (4) | | | | | | | Reset | PRBS 31-bit | PRBS 31-bit |
| Link 0 | MGT_X0Y54/TX | MGT_X0Y54/RX | 27.576 Gbps | 5.732E14 | 0E0 | 1.744E-15 | Reset | PRBS 31-bit | PRBS 31-bit |
| Link 1 | MGT_X0Y55/TX | MGT_X0Y55/RX | 27.575 Gbps | 5.732E14 | 0E0 | 1.744E-15 | Reset | PRBS 31-bit | PRBS 31-bit |
| Link 2 | MGT_X0Y53/TX | MGT_X0Y52/RX | 27.575 Gbps | 5.732E14 | 0E0 | 1.744E-15 | Reset | PRBS 31-bit | PRBS 31-bit |
| Link 3 | MGT_X0Y52/TX | MGT_X0Y53/RX | 27.575 Gbps | 5.732E14 | 0E0 | 1.744E-15 | Reset | PRBS 31-bit | PRBS 31-bit |

Figure 10. Samtec FireFly IBERT loopback test: links status.

A typical eye diagram, obtained using a low power mode of the GTY receiver, is shown in Figure 11. With an open area and an open UI of 9408 and 66.67 % respectively, a very good performance of the Samtec FireFly optical module is achieved.

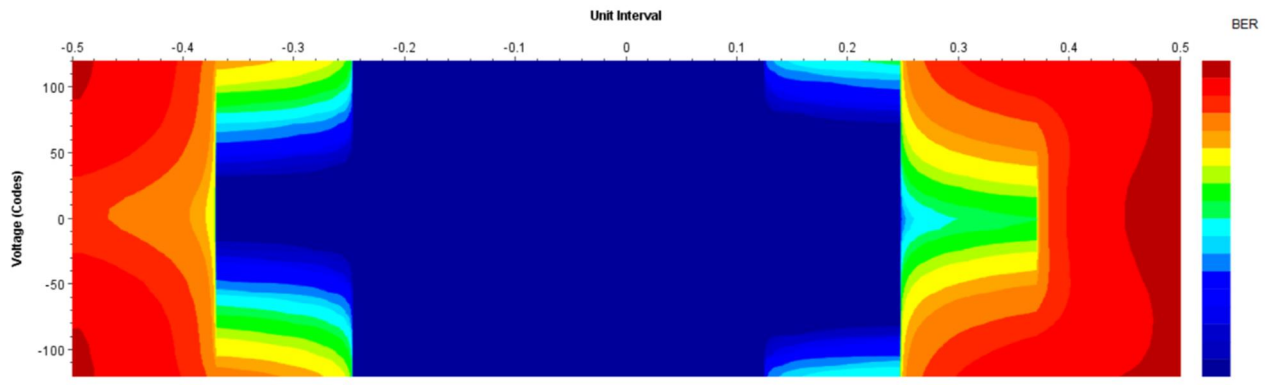

Figure 11. Samtec FireFly IBERT loopback test: a typical eye diagram.

## IV. CONCLUSION

As a part of the R&D for the Phase-II Global Trigger System, a Technological Demonstrator has been designed and tested with a Xilinx Virtex UltraScale+ 9P FPGA and 28 Gb/s Samtec FireFly and Finisar BOA high-speed optical modules implemented on-board. A good performance and absence of bit errors during long runs have been demonstrated. Due to a higher link density and a good performance, Finisar BOA is a good candidate for the Global Trigger and therefore it is used for the PFM.